\documentclass[12pt]{article}
\usepackage{psfig}
\usepackage{verbatim}
\usepackage{color}
\usepackage{rotate}
\usepackage{graphics}
\usepackage{epsf}
\usepackage{epsfig}
\usepackage{graphicx}
\usepackage{latexsym}
\usepackage{amscd}
\usepackage{amssymb}
\newcommand{\beq}{\begin{equation}}
\newcommand{\eeq}{\end{equation}}
\newcommand{\beqn}{\begin{eqnarray}}
\newcommand{\eeqn}{\end{eqnarray}}
\newcommand\noi{\noindent} 
\newcommand\la{\langle}
\newcommand\ra{\rangle}
\newcommand\eps\varepsilon

\def\fm{\,\mbox{fm}}

\def\lsim{\mathrel{\rlap{\lower4pt\hbox{\hskip1pt$\sim$}}
    \raise1pt\hbox{$<$}}}         
\def\gsim{\mathrel{\rlap{\lower4pt\hbox{\hskip1pt$\sim$}}
    \raise1pt\hbox{$>$}}}         

\begin{document}

\hfill {LA-UR-03-5382}

\vspace*{2cm}
\begin{center}
{\Large
\bf 
Coherence length and nuclear shadowing for transverse\\[0.3em] and longitudinal photons}
\end{center}
\vspace{.5cm}

\begin{center}
 {\large
J\"org Raufeisen\footnote{\tt email: jorgr@lanl.gov}}\\
\medskip

{\sl Los Alamos National Laboratory,
Los Alamos, New Mexico 87545, USA}
\end{center}

\vspace{.5cm}

\begin{abstract}
\noi
We study nuclear shadowing for transverse and 
longitudinal photons. The coherence length, which 
controls the onset of nuclear shadowing at small Bjorken-$x$, 
$x_{Bj}$, is longer for longitudinal than for transverse 
photons.  The light-cone Green function technique
properly treats the finite coherence length in all multiple scattering terms.
This is 
especially important in the
region $x_{Bj}>0.01$, where most of the data exist. 
NMC 
data on shadowing in deep inelastic scattering are well reproduced 
in this approach.
We also incorporate nonperturbative effects,
in order to extrapolate this approach to small photon virtualities
$Q^2$, where 
perturbative QCD cannot be applied. This way, we achieve a 
description of shadowing that is based only on quark and gluon 
degrees of freedom, even at low $Q^2$.
\end{abstract}

\vspace*{1.5cm}

\centerline{\em Invited talk given at NAPP03 conference, Dubrovnik, Croatia, 
May 26 -- 31, 2003.}

\clearpage

\section{Introduction}
 
The use of nuclei instead of protons
in high energy scattering experiments,
such as deep inelastic scattering (DIS),
provides unique possibilities to
study the space-time development of strongly interacting systems. In
experiments with proton targets the products of the scattering process can only
be observed in a detector which is separated from the reaction point by a
macroscopic distance. In contrast to this, the nuclear medium can serve as a
detector located directly at the place where the microscopic interaction
happens.
As a consequence, with nuclei one can 
study coherence effects in QCD which are
not accessible in DIS off
protons nor in proton-proton scattering.

\begin{figure}[htb]
\centerline{\psfig{figure=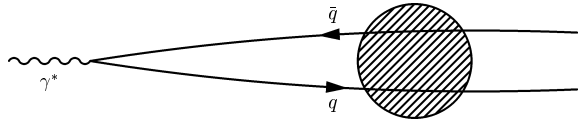,width=8cm}}
\protect\caption{At low $x_{Bj}$ and in the target rest frame, the
virtual photon $\gamma^*$ converts into a $q\bar q$-pair long before
the target.}
\label{fig:disfig}
\end{figure}

At high energies, nuclear scattering is governed by coherence effects
which are most easily understood
in the target rest frame. In the rest frame, 
DIS looks like pair creation from a virtual
photon, see Fig.\ \ref{fig:disfig}. 
Long before the target, the virtual photon splits into a $q\bar
q$-pair. The lifetime $l_c$ of the fluctuation, which is also called
coherence length,
can be estimated
with help of the uncertainty relation to be of order $\sim 1/m_N x_{Bj}$,
where $x_{Bj}$ is Bjorken-x and
$m_N\approx 1$ GeV is the nucleon mass. 
The coherence length can become
much greater than the nuclear radius at low $x_{Bj}$.
Multiple scattering within
the lifetime of the $q\bar q$ fluctuation
leads to the pronounced coherence effects observed in experiment.

The most
prominent example for a coherent interaction of more than one nucleon is the
phenomenon of nuclear shadowing, {\em i.e.} the suppression of 
the nuclear structure function $F_2^A$ with respect to the proton
structure function $F_2^p$, 
$F_2^A(x_{Bj},Q^2)/(AF_2^p(x_{Bj},Q^2))<1$, at low $x_{Bj}\lsim 0.1$. 
Shadowing in low $x_{Bj}$ DIS and at high photon virtualities
is experimentally well studied by NMC \cite{NMC}.

What is the mechanism behind this suppression? 
If the coherence length is very long, as indicated in Fig.~\ref{fig:disfig},
the $q\bar q$-dipole  undergoes multiple scatterings inside the nucleus.
The physics of shadowing in DIS is most easily understood 
in a representation, in which the pair has a
definite transverse size $\rho$. As a result of color transparency
\cite{zkl,bbgg}, small 
pairs interact with a small cross section, while large
pairs interact with a large cross section. 
The term "shadowing" can be taken literally in the target rest frame. Large
pairs are absorbed by the nucleons at the surface which cast a shadow on the
inner nucleons. The small pairs are not shadowed. They have equal chances
to interact with any of the nucleons in the
nucleus. From these simple arguments, one can already understand the two 
necessary conditions for shadowing. First, the hadronic fluctuation of the
virtual photon has to interact with a large cross section and second, the
coherence length has to be long enough to allow for multiple scattering.

\section{Shadowing and diffraction in DIS}

Like shadowing in hadron-nucleus collisions, 
shadowing in DIS is also intimately related to diffraction 
\cite{glaubergribov}. 
The close connection between shadowing and diffraction becomes most transparent
in the formula derived by Karmanov and Kondratyuk \cite{KKK}. In the double
scattering approximation, the shadowing correction can be related to the
diffraction dissociation spectrum, integrated over the mass,
\beq\label{eq:kkk}
{\sigma^{\gamma^*A}}
\approx A\sigma^{\gamma^*p}
-{4\pi}\,
\int dM^2_X\,\left.\frac{d\sigma(\gamma^*N\to XN)}
{dM^2_X\,dt}\right|_{t\to 0}
\times
\int d^2bF_A^2(l_c,b).
\eeq
Here
\beq\label{ff}
F_A^2(l_c,b)=\left|\int_{-\infty}^{\infty}dz\rho_A(b,z)
{\rm e}^{{\rm i}z/l_c}\right|^2
\eeq
is the formfactor of the nucleus, which depends on the coherence length 
\beq\label{length}
l_c=\frac{2\nu}{Q^2+M_X^2},
\eeq
$\nu$ is the energy of the $\gamma^*$ in the target rest frame
and $M_X$ is the mass of the diffractively excited state.
The coherence length can be estimated from the uncertainty relation and 
is the lifetime of the diffractively excited state. If $l_c\to 0$,
the shadowing correction in Eq.~(\ref{eq:kkk}) vanishes and one is left
with the single scattering term $A\sigma^{\gamma^*p}$. 

\subsection{The dipole approach}

Note that Eq.~(\ref{eq:kkk}) is valid only in double scattering approximation.
For heavy nuclei, however, higher order scattering terms will become 
important. These can be calculated, if one knows the eigenstates of the 
interaction. Fortunately, the eigenstates of the $T$ matrix (restricted
to diffractive processes) were identified a long time ago in QCD 
\cite{zkl,mp} as partonic 
configurations with fixed transverse separations in impact parameter space.
For DIS, the lowest eigenstate is the $q\bar q$ Fock component of the
photon. The total $\gamma^*$-proton cross section is easily calculated,
if one knows the cross section $\sigma_{q\bar q}(\rho)$
for scattering a $q\bar q$-dipole of transverse size $\rho$ off a proton,
\beq\label{eq:tot}
\sigma^{\gamma^*p}=\sum_{T,L}\int d\alpha d^2\rho
\left|\Psi^{T,L}_{q\bar q}(\alpha,\rho)\right|^2\sigma_{q\bar q}(\rho).
\eeq
Here, $\alpha$ is the longitudinal momentum fraction carried by the quark
in Fig.~\ref{fig:disfig}.
The light-cone wavefunctions $\Psi^{T,L}_{q\bar q}(\alpha,\rho)$ 
describe the splitting of a  transverse
($T$) and longitudinal ($L$) photon into a $q\bar q$-pair.
For small $\rho$, the light-cone wavefunctions can be calculated in
perturbation theory (see {\em e.g.} \cite{nz} for explicit expressions), 
but at large $\rho$ non-perturbative effects become
important. 
In Ref.~\cite{kst2}, these effects have been modeled by introducing a harmonic 
oscillator potential between the quark and the antiquark, which leads to a
modification of the light-cone wavefunctions. The strength of this potential 
has been determined from data for
photoabsorption on protons. 
This justifies the application of the dipole formulation 
at low $Q^2$ and makes 
this approach an alternative to the vector dominance model (see {\em e.g.}
Ref.~\cite{VDM}).
At large $Q^2$, of course, 
the modified  light-cone wavefunctions reduce to the perturbative ones.

The dipole
cross section is governed by nonperturbative effects and cannot 
be calculated from first principles. 
We use the phenomenological parameterization that is fitted to
HERA data on the proton structure function.
Note that higher Fock-states of the photon, containing gluons, lead
to an energy dependence of $\sigma_{q\bar q}$, which we do not write
out explicitly.

The diffractive cross section can also be expressed in terms of
$\sigma_{q\bar q}$. Since the cross section for diffraction is
proportional to the square of the $T$-matrix element, 
$|\la\gamma^*|T|X\ra|^2$, the dipole cross section also
enters squared,
\beq\label{eq:diff}
\int dM^2_X\,\left.\frac{d\sigma(\gamma^*N\to XN)}
{dM^2_X\,dt}\right|_{t\to 0}=\frac{\la\sigma^2_{q\bar q}(\rho)\ra}
{16\pi},
\eeq
where the brackets $\la\dots\ra$ indicate averaging over the light-cone
wavefunctions like in Eq.~(\ref{eq:tot}). We point out that in
order to reproduce the
correct behavior of the diffractive cross section at large $M_X$, one has to
include at least the $q\bar qG$ Fock-state of the $\gamma^*$. This correction 
is, however, of minor importance in the region where shadowing data are 
available.  

\subsection{The Green function technique}

If one attempts to calculate shadowing from Eq.~(\ref{eq:kkk}) with help of
Eq.~(\ref{eq:diff}), one faces the problem that the nuclear form factor, 
Eq.~(\ref{ff}), depends on the mass $M_X$ of the diffractively produced
state, which is undefined in impact parameter representation.
Only in the limit
$l_c\gg R_A$, where $R_A$ is the nuclear radius,
it is possible to resum the entire multiple scattering series
in an eikonal-formula
\beq\label{eikonalappr}
\sigma^{\gamma^*A}=\left< 2\int d^2b\left(
1-\exp\left(-\frac{\sigma_{q\bar q}(\rho)}{2}T(b)\right)\right)\right>.
\eeq
The nuclear thickness function 
$T(b)=\int_{-\infty}^{\infty}dz\,\rho_A(b,z)$ is the
integral of
nuclear density over longitudinal
coordinate $z$ and depends on the
impact parameter $b$.
The condition $l_c\gg R_A$
makes sure that the $\rho$ does not vary during propagation
through the nucleus (Lorentz time dilation) and one can apply the eikonal
approximation.

\begin{figure}[htb]
\centerline{\psfig{figure=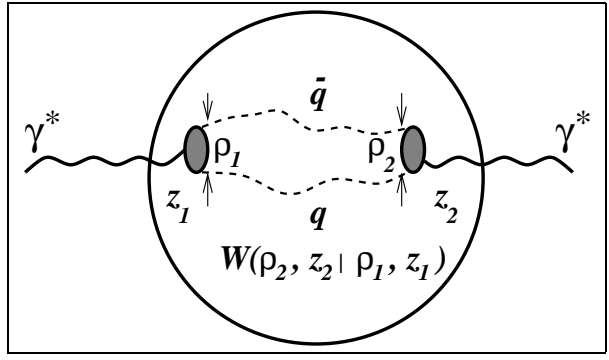,width=8cm}}
\protect\caption{Propagation of a $q\bar q$-pair through a nucleus. 
Shown is the case of
a finite coherence length, where the transverse motion is described by the
Green function $W\left(\vec\rho_2,z_2|\vec\rho_1,z_1\right)$. }
\label{fig:propag}
\end{figure}

The condition $l_c\gg R_A$ is however not fulfilled in experiment.
For the case $l_c \sim R_A$, one has to take the
variation of
$\rho$ during propagation of the $q\bar q$ fluctuation through
the nucleus into account, see Fig.~\ref{fig:propag}. A widely used recipe is
to replace $M_X^2\to Q^2$, so that $l_c\to 1/(2m_Nx_{Bj})$ and one
can apply the double scattering approximation. This recipe was, however,
disfavored by our investigation \cite{krt1}. Moreover, there
is no simple recipe to include a finite $l_c$ into higher order scattering 
terms.

In Ref.~\cite{krt1} (see also Ref.~\cite{urs})
a Green function technique was developed that 
provides the correct quantum-mechanical treatment of a finite coherence length
in all multiple scattering terms.
Like in Eq.~(\ref{eq:kkk}) the total cross section 
is represented in the form
\beq\label{form}
\sigma^{\gamma^*A}=A\sigma^{\gamma^*p}-\Delta\sigma,
\eeq
where $\Delta\sigma$ is the shadowing correction,
\beqn\nonumber\label{correction}
\Delta\sigma&=&\frac{1}{2}{\rm Re}\sum_{T,L}\int d^2b
\int\limits_{-\infty}^{\infty} dz_1\rho_A(b,z_1)
\int\limits_{z_1}^{\infty} dz_2\rho_A(b,z_2)\\
&\times&\nonumber
\int_0^1d\alpha\int d^2\rho_2
\left[\Psi_{q\bar q}^{T,L}(\vec\rho_2,\alpha)\right]^*\sigma_{q\bar q}(\rho_2)
A(\vec\rho_2,z_1,z_2,\alpha),
\eeqn
with
\beq\label{propagation}
A(\vec\rho_2,z_1,z_2,\alpha)=\int\,d^2\rho_1\,
W(\vec \rho_2,z_2|\vec \rho_1,z_1)\,
{\rm e}^{-{\rm i}q_L^{min}(z_2-z_1)}\,\sigma_{q\bar q}(\rho_1)\,
\Psi^{T,L}_{q\bar q}(\vec \rho_1,\alpha).
\eeq
Here,
\beq
q_L^{min}=\frac{1}{l_c^{max}}=\frac{Q^2\alpha(1-\alpha)
+m_f^2}{2\nu\alpha(1-\alpha)}
\eeq
is the minimal longitudinal momentum transfer when the photon splits into the
$q\bar q$ dipole ($m_f$ is the quark mass). 

The shadowing term in Eq.~(\ref{form}) is
illustrated
in Fig.~\ref{fig:propag}.  
At the point $z_1$ the photon diffractively produces the $q\bar q$
pair ($\gamma^*N\to q\bar qN$) with a transverse separation $\vec\rho_1$.
The pair propagates through the nucleus along arbitrarily curved
trajectories, which are summed over, and arrives at the point
$z_2$ with a separation $\vec\rho_2$.  The initial and the
final separations are controlled by the light-cone wavefunctions 
$\Psi^{T,L}_{q\bar q}(\vec \rho,\alpha)$.  While passing the nucleus
the $q\bar q$ pair interacts with bound nucleons
via the cross section $\sigma_{q\bar q}(\rho)$ which depends on the local
separation $\vec\rho$.  The Green function 
$W(\vec\rho_2,z_2|\vec\rho_1,z_1)$ describes the propagation of the pair from
$z_1$ to $z_2$, see Eq.~(\ref{propagation}), including all
multiple rescatterings and a finite coherence length.
Note the diffraction dissociation ($DD$) amplitude,
\beq\label{diffamp}
f_{DD}(\gamma^*\to q\bar q)={\rm i}\Psi^{T,L}_{q\bar q}(\vec \rho_1,\alpha)
\sigma_{q\bar q}(\rho_1),
\eeq
in Eq.~(\ref{propagation}).
At the position $z_2$, the result of the propagation is again 
projected onto the
diffraction dissociation amplitude. 
The Green function
includes that part of the phase shift between the
initial and the final photons which is due to transverse
motion of the quarks, while the longitudinal motion is included in
Eq.~(\ref{propagation}) through the exponential.

\begin{figure}[t]
\centerline{\psfig{figure=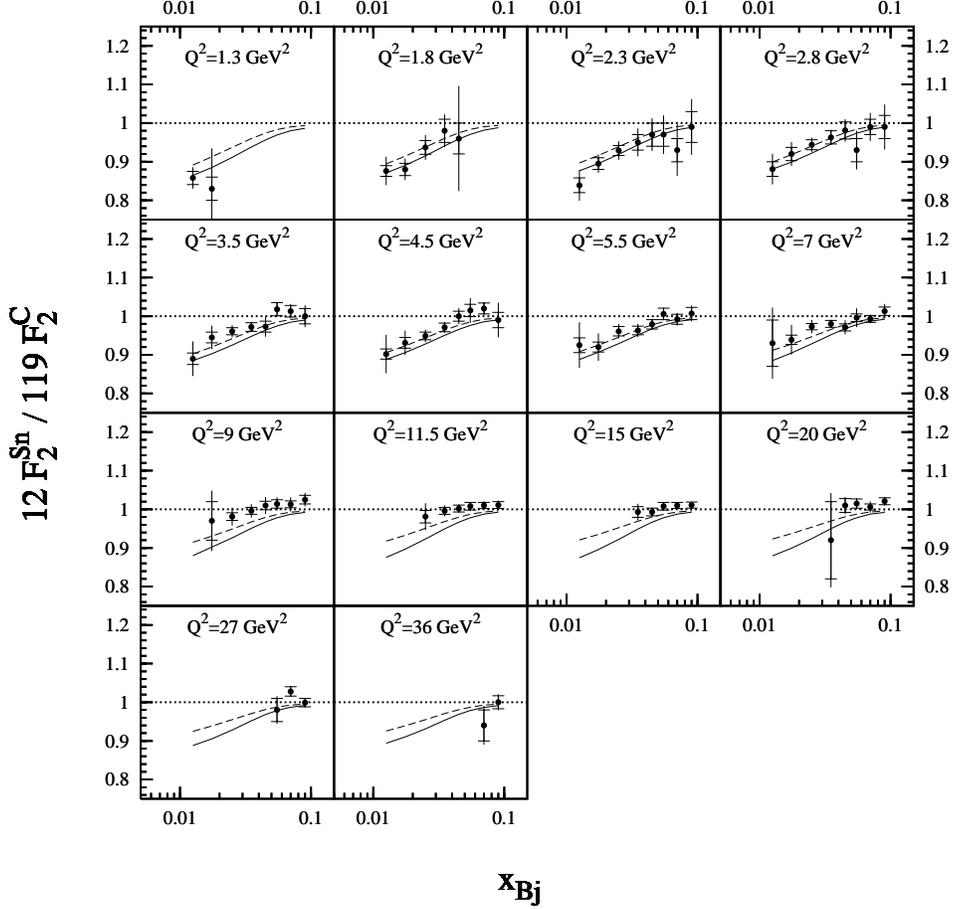,width=13cm}}
\protect\caption{The $x_{Bj}$ dependence of shadowing in DIS
for the structure function of tin relative to carbon. The data are 
from NMC \cite{NMC}. The full curves are calculated in the Green-function
approach, including the nonperturbative interaction between the $q$ and the 
$\bar q$. The dashed curve does not include this interaction. The figure is 
from Ref.~\cite{thesis}. }
\label{fig:nmc}
\end{figure}

The Green function $W(\vec \rho_2,z_2;\vec\rho_1,z_1)$ in
Eq.~(\ref{propagation}) satisfies the two dimensional 
Schr\"o\-din\-ger equation,
\beq\label{schroedinger}
{
{\rm i}\,\frac{{\partial}W(\vec\rho_2,z_2|\vec\rho_1,z_1)}{{\partial}z_2}
=}
-\frac{\Delta(\rho_2)}{2\nu\alpha(1-\alpha)}\,
W(\vec\rho_2,z_2|\vec\rho_1,z_1)
-{{\rm i}\over2}\,\sigma(\rho_2)\,\rho_A(b,z_2)\,
 W(\vec\rho_2,z_2|\vec\rho_1,z_1)
\eeq
with the boundary condition $W(\vec\rho_2,z_1|\vec\rho_1,z_1)
=\delta^{(2)}(\vec \rho_2-\vec \rho_1)$.  The Laplacian $\Delta(\rho_2)$ acts
on the coordinate $\vec\rho_2$.   
The kinetic term $\Delta/[2\nu\alpha(1-\alpha)]$ in this Schr\"odinger equation
takes care of the varying effective mass of the $q\bar q$ pair
and provides the proper phase shift.
The role of time is played by the
longitudinal coordinate $z_2$.
The imaginary part of the optical potential describes the rescattering.
This equation has recently been solved numerically by Nemchik~\cite{nemchik}.

The Green function method contains  the Karmanov-Kondratyuk formula 
Eq.~(\ref{eq:kkk}) and the eikonal
approximation Eq.~(\ref{eikonalappr}) as limiting cases.
In order to obtain the eikonal approximation, one has to take the limit
$\nu \to\infty$. In this case, the kinetic energy term in
Eq.~(\ref{schroedinger}) can be neglected and  
with $q_L^{min}\to 0$ one arrives 
after a short calculation at  Eq.~(\ref{eikonalappr}).
One can also recover the Karmanov-Kondratyuk formula, 
when one neglects 
the imaginary potential in Eq.~(\ref{schroedinger}). 
Then $W$ becomes the  Green function of a free motion.

Calculations in the Green function approach are compared to NMC data in
Fig.~\ref{fig:nmc}. Note that the leading twist contribution to shadowing 
is due to large dipole sizes, where nonperturbative effects, such as
an interaction between the $q$ and the $\bar q$, might become important
(see above).
Therefore, the solid curve is calculated with the modified light-cone
wavefunctions of
Ref.~\cite{kst2}. 
The dashed curve is calculated with the conventional, perturbative light-cone
wavefunctions, but including a constituent quark mass. Both curves are 
in reasonable agreement with the data. We stress that the Green function 
technique also takes into account some higher twist corrections to shadowing.
This is essential for a successful description of the NMC data. In fact,
it has been demonstrated in Ref.~\cite{guzey} that the leading twist 
approximation only poorly reproduces NMC data for shadowing in calcium.

Note that 
for the data shown in Fig.~\ref{fig:nmc}, 
the coherence length is of order of the
nuclear radius or smaller. Indeed, shadowing vanishes around 
$x_{Bj}\approx 0.1$, because the coherence length becomes smaller than
the mean internucleon spacing. Therefore, the eikonal approximation, 
Eq.~(\ref{eikonalappr}), cannot be applied for the kinematics of NMC
and a correct treatment of the coherence length becomes crucial.
We emphasize that the calculation in Fig.~\ref{fig:nmc} does not 
contain any free parameters. Following the spirit of Glauber theory,
all free parameters are adjusted to DIS off protons.

\subsection{Higher Fock states and the leading twist gluon shadowing}

The shadowing for quarks discussed in the previous subsection is
dominated by the transverse photon polarization. 
Longitudinal photons, on the other hand,
 can serve to measure the gluon density because they
effectively couple to color-octet-octet dipoles. This can be understood in
the following way: the light-cone wave function for the transition
$\gamma^*_L\to q\bar q$ does not allow for large, aligned jet configurations
as is the case for transversely polarized photons. 
Thus, all $q\bar q$ dipoles from longitudinal
photons have size $1/Q^2$ and the double-scattering term vanishes
$\propto 1/Q^4$. The leading-twist contribution for the shadowing of 
longitudinal
photons arises from rescattering of
the $|q\bar q G\ra$ Fock state of the photon. Here again,
the distance between the $q$ and the $\bar q$ is of order $1/Q^2$, but the
gluon can propagate relatively far from the $q\bar q$-pair. In addition,
after radiation of the gluon, the pair is in an octet state. Therefore, the
entire $|q\bar qG\ra$-system appears as a $GG$-dipole, and the shadowing
correction to the longitudinal cross section can be identified with
gluon shadowing.

A critical issue for determining the magnitude of gluon shadowing is the 
distance the gluon can propagate from the $q\bar q$-pair in impact parameter 
space, {\em i.e.} knowing how large the $GG$ dipole can become.  
This value has been extracted from single diffraction data in hadronic 
collisions in Ref.~\cite{kst2} because these data allow the 
diffractive gluon radiation (the triple-Pomeron contribution in Regge
phenomenology) to be unambiguously singled out.
The diffraction cross section ($\propto \rho^4$) is even
more sensitive to the dipole size than the total cross section ($\propto
\rho^2$) and is therefore a sensitive probe of the mean transverse
separation. It was found in Ref.~\cite{kst2} that the mean dipole size
must be of the order of $r_0=0.3\,\fm$, considerably smaller than a light 
hadron. A rather small gluon cloud of this size surrounding the valence 
quarks is the only way that is known to resolve the long-standing problem of 
the small size of the triple-Pomeron coupling.  
The smallness of the $GG$ dipole is incorporated into the LC approach by 
a nonperturbative interaction between the gluons. 

\begin{figure}[t]
\centerline{\hfill\psfig{figure=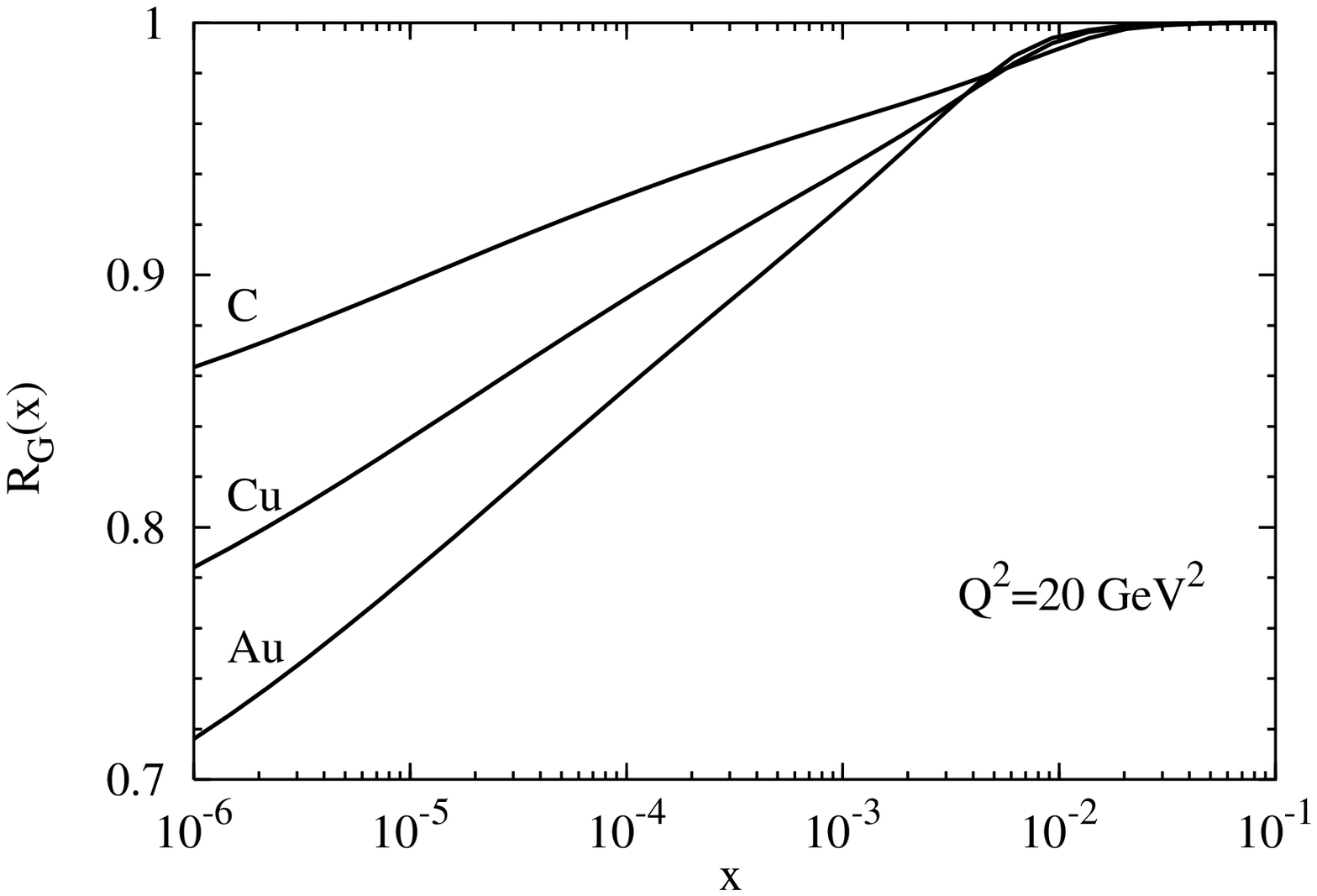,width=7cm}\hfill\psfig{figure=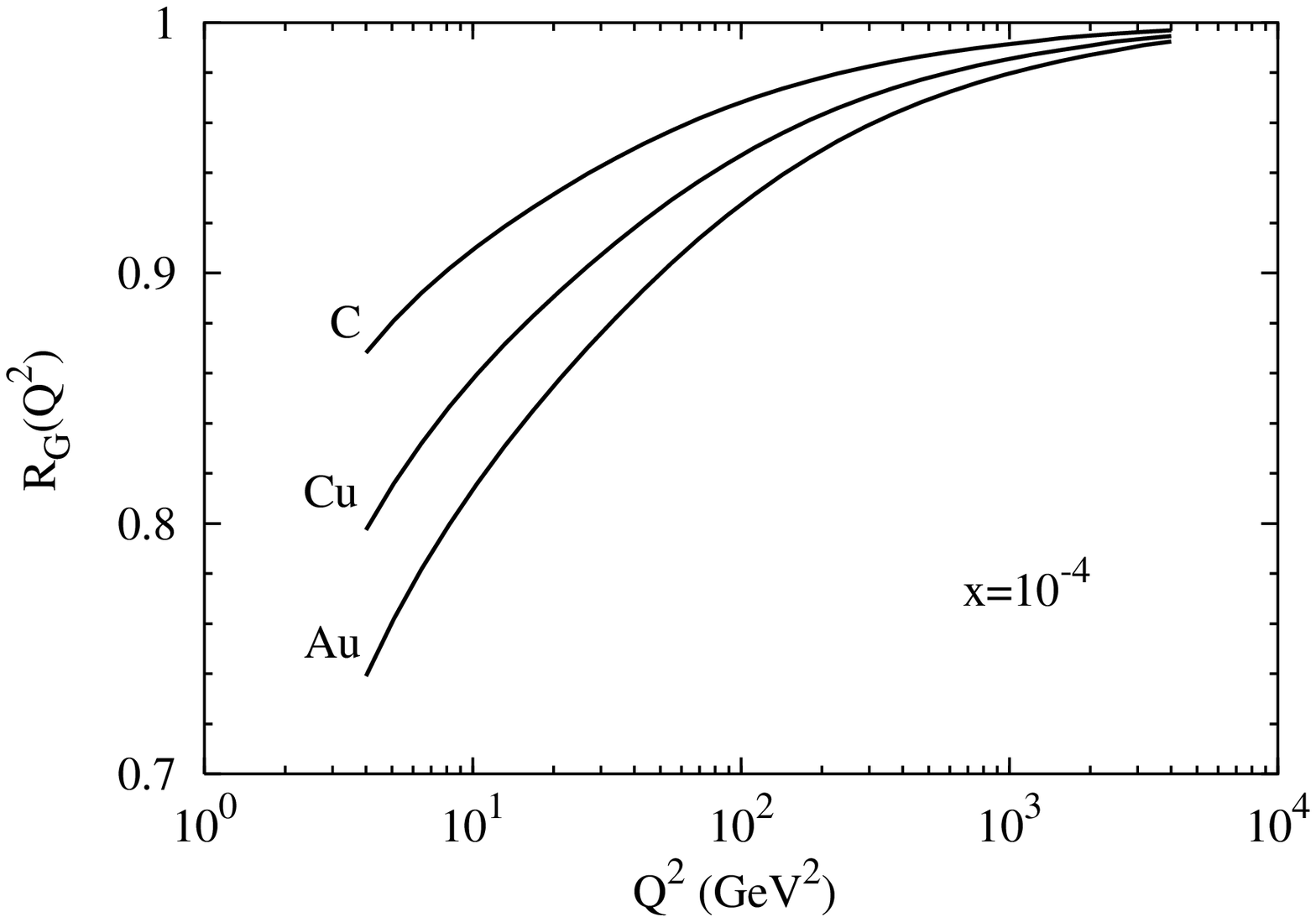,width=7cm}\hfill}
\protect\caption{The $x$- and $Q^2$-dependence of gluon shadowing for carbon, copper and
gold. The $x$-dependence is shown for
$Q^2=20$ GeV$^2$, while the figure on the right is calculated for
$x=10^{-4}$.  }
\label{fig:gshad}
\end{figure}

We calculate gluon shadowing as function of $x$ at fixed $Q^2$ and as
a function of $Q^2$ at fixed $x$, integrated over the impact parameter $b$. The
results are shown in Fig.\ \ref{fig:gshad}.
In the left-hand 
plot, one observes that gluon shadowing vanishes for $x>0.01$. This
happens because the lifetime of the $|q\bar qG\ra$-fluctuation becomes smaller
than the mean internucleon distance of $\sim 2$~fm as $x$ exceeds $0.01$.
Indeed, in Ref.~\cite{krt2} an average coherence length of slightly less than 
$2$~fm was found for the $|q\bar qG\ra$-state at $x=0.01$ and large $Q^2\gg
1/r_0^2$. Note that gluon shadowing sets in at a smaller value of $x$ than
quark shadowing because the mass of a $|q\bar q G\ra$-state is larger than
the mass of a $|q\bar q\ra$-state.   We also point out that gluon shadowing
is weaker than quark shadowing in the $x$-range plotted, because the
small size of the $GG$-dipole overcompensates the Casimir factor in the
$GG$-proton cross section, $\sigma_{GG}=(9/4)\sigma_{q\bar q}$.  
Note however that at this time, almost nothing is known about gluon 
shadowing from experiment, and theoretical approaches differ vastly
(see {\em e.g.} Ref.~\cite{nestor} for a comparison of different models). 
The plot
on the right-hand side 
of Fig.\ \ref{fig:gshad} shows the $Q^2$-dependence of gluon
shadowing and clearly demonstrates that gluon shadowing is a leading-twist
effect, with $R_G$ only very slowly (logarithmically) approaching unity as
$Q^2\to\infty$. 
 
\subsection{The mean coherence length}

The importance of the coherence length $l_c$ was already mentioned above:
this quantity controls the onset of quark and gluon shadowing at small
$x_{Bj}$ and therefore governs shadowing in the kinematical region
that is of interest for most experiments. 
The Green function technique is the only known way to include
the quantum mechanically correct coherence length into all 
multiple scattering terms. However, since $l_c$ is not well defined in
impact parameter space, it appears only implicitly in the Green function 
approach. It is therefore useful to introduce the concept of a mean 
coherence length, as it was done in Ref.~\cite{krt2}.

A photon of virtuality $Q^2$ and energy $\nu$
can develop a hadronic fluctuation for a lifetime,
\beq
l_c=\frac{2\,\nu}{Q^2+M_{q\bar q}^2}=
\frac{P}{x_{Bj}\,m_N}\ ,
\label{1.1}
\eeq
where 
$M_{q\bar q}$ is the effective mass of the fluctuation,
and the factor $P^{-1}=(1+M_{q\bar q}^2/Q^2)$. 
The usual approximation is to assume that
$M_{q\bar q}^2 \approx Q^2$ since $Q^2$ is the only large 
dimensional scale available. In this case $P=1/2$.

The effective mass of a non-interacting $q\bar q$-pair 
is well defined, $M_{q\bar q}^2=(m_f^2+p_T^2)/\alpha(1-\alpha)$,
where $p_T$ and $\alpha$ are the transverse
momentum and fraction of the
light-cone momentum of the photon carried by the quark, respectively.
Therefore, $P$ has a simple form,
\beq
P(k_T,\alpha)=\frac{Q^2\,\alpha\,
(1-\alpha)}{p_T^2+\eps^2}\ ,
\label{1.2}
\eeq
where
$
\eps^2 = \alpha(1-\alpha)Q^2 + m_f^2
$.

To find the mean  lifetime of those fluctuations that contribute to shadowing,
one should define the averaging procedure as
\beq
\la P\ra=
\frac{\Bigl\la f(\gamma^*\to q\bar q)
\Bigl|P(p_T,\alpha)\Bigr|
f(\gamma^*\to q\bar q)\Bigr\ra}
{\Bigl\la f(\gamma^*\to q\bar q)\Bigl|
f(\gamma^*\to q\bar q)\Bigr\ra}\ ,
\label{1.6a}
\eeq
where $f(\gamma^*\to q\bar q)$ is the amplitude of diffractive
dissociation of the virtual photon on a nucleon, Eq.~(\ref{diffamp}).
That way, $P$ is weighted with the 
interaction cross section squared $\sigma^2_{q\bar q}(\rho)$
in the averaging procedure.
Then, the mean value of the factor $P(\alpha,p_T)$ reads for transverse
and longitudinal photons,
\beq
\left\la P^{T,L}\right\ra=\frac{\int_0^1d\alpha\int\! d^2\rho_1d^2\rho_2
\left[\Psi_{q\bar q}^{T,L}\left(\vec \rho_2,\alpha\right)\right]^*\!\!
\sigma_{q\bar q}\left(\rho_2\right)
\widetilde P\left(\vec \rho_2-\vec \rho_1,\alpha\right)
\Psi_{q\bar q}^{T,L}\left(\vec \rho_1,\alpha\right)
\sigma_{q\bar q}\left(\rho_1\right)}
{\int_0^1d\alpha\int d^2\rho
\left|\Psi_{q\bar q}^{T,L}\left(\vec \rho_,\alpha\right)
\sigma_{q\bar q}\left(\rho\right)\right|^2}
\label{1.7}
\eeq
with
\beq
\widetilde P\left(\vec \rho_2-\vec \rho_1,\alpha\right)=
\int\frac{d^2p_T}{\left(2\pi\right)^2}\,
{\exp\left(-{\rm i}\,\vec p_T\cdot\left( \vec \rho_2-\vec \rho_1\right)\right)}
{P\left(\alpha,\rho\right)}.
\label{1.8}
\eeq

\begin{figure}[t]
\centerline{\psfig{figure=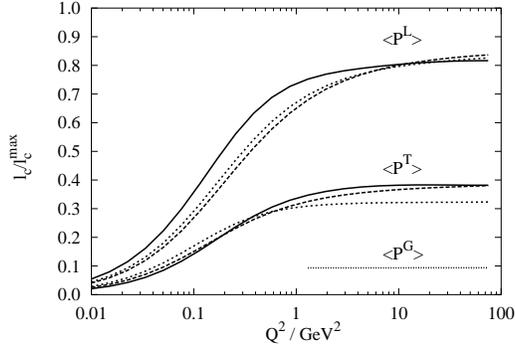,width=7cm}}
\protect\caption{The mean coherence length as function of $Q^2$. 
The curves are
        for $q\bar q$ fluctuations of transverse 
        and longitudinal
        photons, and for $q\bar qG$ fluctuation, from the top to 
        bottom, respectively. Dot\-ted curves correspond to
        calculations with perturbative wavefunctions 
        and an approximate dipole
        cross section $\propto \rho^2$. Dashed curves are the same, except
        the realistic parameterization of $\sigma_{q\bar q}(\rho)$ is 
	employed (see Ref.~\cite{krt2} for details).
        The solid curves are calculated with 
        the nonperturbative wavefunctions of Ref.~\cite{kst2}  }
\label{fig:meanl}
\end{figure}

Numerical results are shown in Fig.~\ref{fig:meanl}. 
Note that the nonperturbative interaction also modifies the invariant mass
of the $q\bar q$ pair from Eq.~(\ref{1.2}). This was taken into account in
Ref.~\cite{krt2}.
One observes that the
mean coherence length for longitudinal photons is approximately 
twice as long as for transverse photons. However, shadowing for the $q\bar q$
Fock component of a longitudinally polarized photon is higher twist (see 
above). The coherence length for the $q\bar q G$ Fock state of a longitudinal
photon, which gives rise to leading twist gluon shadowing, is much shorter, 
resulting in a delayed onset of shadowing for gluons. 

\section{Summary}

DIS at low $x_{Bj}$ is most naturally described in the color dipole
formulation, because partonic configurations with fixed transverse 
separations in impact parameter space are eigenstates of the interaction.
This salient feature allows one to calculate multiple scattering effects,
such as nuclear shadowing, in a very easy way: at very high energies, 
one can simply eikonalize the dipole cross section. At realistic energies,
however, corrections due to the finite lifetime $l_c$ of the $q\bar q$-pair
become important. In Ref.~\cite{krt1}, we succeeded in 
generalizing the Gauber-Gribov theory of nuclear shadowing by
incorporating a
finite $l_c$ into all multiple scattering terms, using a Green function 
technique.  
Since nuclear shadowing is dominated by large dipole sizes, a nonperturbative
interaction in form of an harmonic oscillator potential between the quarks
was introduced in Ref.~\cite{kst2}. This makes the dipole approach 
applicable at low $Q^2$. Note however that it is not necessary for
this model to  reproduce the vector meson masses or the coherence length 
of the vector dominance model \cite{krt2}.

The main nonperturbative input to all formulae, the dipole cross section,
cannot be calculated from first principles. Instead we use a
phenomenological parameterization for this quantity, which is determined
from low $x_{Bj}$ DIS. In the spirit of Glauber theory,
nuclear effects are then calculated without 
introducing any new parameters.
This way a good description of NMC
data on shadowing in DIS is achieved. 
We did not attempt to include antishadowing, since this effect probably
is beyond the standard shadowing dynamics.

The parameterization of the dipole cross section   effectively also 
includes higher Fock states (containing gluons) through its energy 
dependence. These higher Fock states are however
excluded from nuclear effects,
if one eikonalizes only the $q\bar q$ cross section. The most striking
consequence is that gluon shadowing ({\em i.e.} shadowing for longitudinal
photons) appears to be higher twist. This problem is overcome
by calculating the rescattering of the $|q\bar qG\ra$-Fock state of the virtual
photon, which gives rise to the leading twist gluon shadowing. 
A detailed calculation is published in Refs.~\cite{kst2,krtj}.
We emphasize that gluon shadowing sets in at smaller $x_{Bj}$ than 
shadowing for quarks, because the larger mass of the $|q\bar qG\ra$-Fock state
leads to a shorter coherence length. This result is supported by
a calculation of the mean coherence length \cite{krt2}. Shadowing disappears,
when the coherence length becomes shorter than the mean internucleon 
separation ($\sim 2\fm$). 

The main advantage of the dipole formulation and our motivation for pursuing
this approach is the insight it provides into the dynamical origin of nuclear
effects, which are calculated without free parameters.
Note that a variety of other processes can be described in the dipole 
language, {\em e.g.} Drell-Yan (DY)
dilepton production \cite{rpn}, gluon radiation
\cite{kst1}
heavy quark \cite{npz} and quarkonium production \cite{kr}.
A parameter free calculation of nuclear 
shadowing for DY for example is needed, 
if one aims at extracting the energy loss
of a fast quark propagating through nuclear from DY data \cite{eloss}.
Furthermore, this approach can also be applied to
calculate the Cronin effect at RHIC \cite{ce}.

Finally, we point out that even though the mechanism of a hard reaction
looks quite differently in the dipole formulation from what one is
used to in the parton model, the 
equivalence between the dipole approach and the conventional parton model
has been demonstrated (for single scattering)
numerically \cite{rpn} and analytically \cite{rp}
for several processes.

\medskip
\noindent {\bf Acknowledgments}: 
I am grateful to the organizers of NAPP03 conference for inviting me to this
stimulating meeting.
This work was supported by the
U.S.~Department of Energy at Los Alamos National Laboratory under Contract
No.~W-7405-ENG-38.

\end{document}